\documentclass[manuscript]{rmaa}
\usepackage{graphicx}
\usepackage{natbib}
\title{IMF from infrared photometry of young stellar clusters in Taurus-Auriga and Orion}

\author{Luis Salas
and Irene Cruz-Gonz\'alez \affil{Instituto de Astronom\'{\i}a, Universidad Nacional Aut\'onoma de M\'exico}}

\shortauthor{Salas, \& Cruz-Gonz\'alez}
\shorttitle{YSC masses from JHKL photometry}

\resumen{Se aplic\'o el m\'etodo de vectores principales de disco y extinci\'on a los datos fotom\'etricos en el cercano infrarrojo de los c\'umulos estelares j\'ovenes de las regiones Tauro-Auriga y Ori\'on. Bajo la suposici\'on que la edad del c\'umulo est\'a representada por la mediana de la distribuci\'on de edades se obtiene una estimaci\'on de las masas estelares individuales. La funci\'on inicial de masa (FIM) obtenida para estos c\'umulos exhibe una gran similitud con la obtenida por m\'etodos espectrosc\'opicos y fotom\'etricos y puede constituir una representaci\'on robusta de la FIM. El m\'etodo permite encontrar la contribuci\'on por  extinci\'on y disco para cada estrella. La distribuci\'on global de la extinci\'on para el c\'umulo de Ori\'on se compara bi\'en con resultados previos. Se encuentra un dominio en la frecuencia de estrellas T Tauri con disco.}

\abstract{We applied the extinction-disk-principal vectors approach to near infrared photometric data of  the Taurus-Auriga region and Orion Nebula young stellar clusters. By assuming that the cluster age is represented by the median value of the age distribution we are able to derive the distribution of stellar masses.  We showed that the resulting initial mass function (IMF) for these two young stellar clusters  compares remarkably well and might be a robust representation of the IMF obtained by spectroscopic or photometric methods. The method also yields extinction and disk contribution for each star. The overall extinction distribution for the Orion cluster is analyzed and compares well with previous work. The frequency of T Tauri stars with disks is dominant.}

\addkeyword{Stars: formation}
\addkeyword {Stars: fundamental parameters} 
\addkeyword{Stars: low-mass} 
\addkeyword{Stars: pre-main sequence}

\fulladdresses{
\item Irene Cruz-Gonz\'alez: Instituto de
Astronom\'{\i}a, Universidad Nacional Aut\'onoma de M\'exico, Ciudad
Universitaria, Apartado Postal 70--264,  04510, M\'exico D.F.,
M\'exico (irene@astroscu.unam.mx).
\item  Luis Salas: Instituto de Astronom\'{\i}a, Universidad Nacional Aut\'onoma de M\'exico (salas@astrosen.unam.mx).
Apdo. Postal 877,  Ensenada, B.~C. 22830, M\'exico.
}

\listofauthors{L.~Salas, \& I.~Cruz-Gonz\'alez}
\indexauthor{Salas,~L.}
\indexauthor{Cruz-Gonz\'alez,~I.}

\begin{document}
\maketitle 
 
\section{Introduction}

Masses of pre-main sequence (PMS) objects are quite difficult to determine, and the only direct way to assess them is through the analysis of their dynamical parameters.  Following \citet{1979ApJS...41..743C}  locating objects  in a  temperature-luminosity H-R diagram along with evolutionary models yields masses.  Since stellar temperature  is estimated from colors, and these are affected by circumstellar disks, PMS temperatures are best estimated spectroscopically \citep[e.g.,][]{2005ApJ...618..810L}. But PMS objects are usually faint and found in dusty environments and obtaining their spectra becomes quite difficult. This has lead many authors to explore the initial mass function (IMF) of young stellar clusters in the infrared, via the luminosity function in the $K$-band \citep[e.g.,][]{2002ApJ...573..366M}, assuming that it reproduces the true IMF of a cluster. Color-color (CC)  and color-magnitude (CM) near-infrared diagrams of young stellar clusters show that intrinsic near-infrared colors of young stars are affected both by interstellar extinction and by disk excess emission  \citep{1992ApJ...393..278L,1992ApJ...397..613H,1997AJ....114..288M,
2000ApJ...540..236H}.

\citet{2007RMxAA..43..155L}, hereon {LS07}, showed that masses of T Tauri stars can be obtained using $JHK$ photometry and both CC and CM diagrams.  This is the result of the analysis of two principal vectors, one produced by the disk excess, $\vec{D}$ and the other by interstellar extinction, $\vec{X}$, if stellar ages are known and a particular set of evolutionary tracks is assumed.

In this paper we continue the development of a new approach in determining masses of pre-main sequence stars from near-infrared photometry.  Our goal is to strengthen the results reported previously in LS07. First, we refine the values of disk excess coefficients given in LS07 by showing that the method can be extended to the $I$ and $L$ filters, and that these coefficients scale well with wavelength. This analysis is presented in section~2. Second, we test whether the method can be used to extract the Initial Mass Function (IMF) of young stellar clusters from infrared photometric data alone. Sect.~3 shows that if the age of a young cluster is known the distribution of stellar masses can be obtained from their $JHK$ photometry and a set of PMS evolutionary tracks.  As a proof of this statement we apply the method to the well studied  {Taurus-Auriga} region and the  {Orion Nebula Cluster}. It is shown that the median age of the clusters produces an excellent agreement with previously known IMFs. These results  are reported in Sects.~4 and 5, while a summary of our results is presented in Sect.~6.

\section{Method description}

\citet{2007RMxAA..43..155L} showed that masses of T Tauri stars can be obtained using their $JHK$ photometry and their location in both CC  and  CM diagrams.  This is the result of the analysis of two principal vectors, one produced by the disk contribution, $\vec{D}$, and the other by interstellar extinction, $\vec{X}$.  LS07 show that the vector $\widehat{X}$ can be defined as the extinction vector corresponding to $A_V$=10 with components given  in  \citet{1985ApJ...288..618R}. While  the vector $\widehat{D}$ is obtained via $J$, $H$,  and $K$ magnitudes obtained from \citet{2005RMxAA..41...61D} models of accretion disks  irradiated by a central T Tauri star.  These  base vectors were thus given the following components in CM ( $K$ vs.  $(J-K)$) and CC ($(J-H)$ vs. $(H-K)$) diagrams:
\begin{eqnarray}
 \widehat{D_\mathrm{cm}} = \left(1.014,-1.105\right)  \,;\,  \widehat{X_\mathrm{cm}} = \left( 1.68,1.16\right)  \nonumber \\
\\
\widehat{D_\mathrm{cc}}  = \left(0.767,0.221\right)    \,;\,  \widehat{X_{cc}} = \left(0.61,1.07\right) \nonumber
\end{eqnarray}
The position of a particular star in these CC and CM diagrams becomes then a vector sum in each diagram that can be reduced to a closure relation plus the following set of linear equations:
\begin{eqnarray}
J \,=\,  J_0 \,+\, \alpha x_\mathrm{J} \,-\,  \beta y_\mathrm{J} \nonumber \\
H \,=\,  H_0 \,+\, \alpha x_\mathrm{H} \,-\,  \beta y_\mathrm{H}  \\
K \,=\,  K_0 \,+\, \alpha x_\mathrm{K} \,-\,  \beta y_\mathrm{K} \nonumber 
\end{eqnarray}
where $\alpha$ and $\beta$ represent the amount of extinction and infrared excess contributions to the individual magnitudes, $x_\lambda$ and $y_\lambda$ are the corresponding extinction and disk coefficients for each wavelength $\lambda$, and the $0$ sub-indices denote the intrinsic photospheric magnitudes.  Values for  $x_\lambda$ and $y_\lambda$ are obtained directly from
the components of base vectors given in eq. (1) and are presented in
Table~1.

\begin{figure}
\includegraphics[width=8cm]{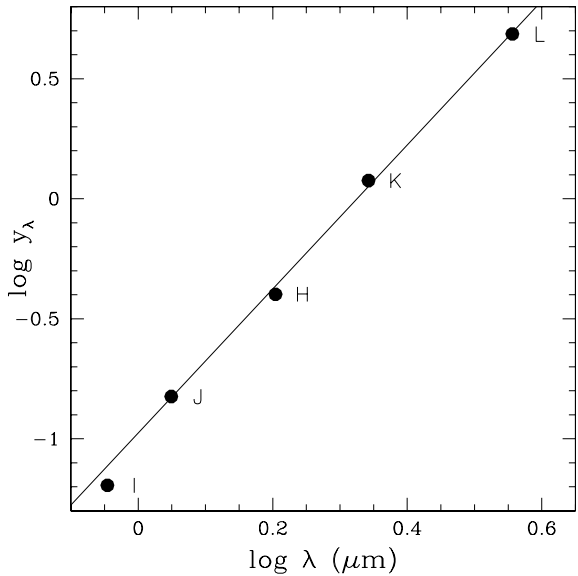}
\caption{ Plot of $\log y_{\lambda}$ as a function of $\log \lambda$ shows that for IJHKL the data are well represented by a  power-law of index=3, shown as a solid line.}
\label{fig1}
\end{figure}

An advantage of presenting the equations as a set of linear equations is that it can be easily explored with any set of three photometric filters at a time, that is, we may use filters $IJK$ or $JKL$ instead of $JHK$, provided the corresponding closure condition is fulfilled and that the excess colors behave as vectors. We found that this procedure is also feasible for filters $I$ and $L$ in addition to the $JHK$ set, and so the numerical values of the  $y_\lambda$  coefficients for $I$ and $L$ are included in Table~1. We show in Figure~\ref{fig1} that the relation between  $\log y_\lambda$ and  $\log \lambda$ can be well represented by a power-law index equal to 3. This close relation gives us confidence to refine the $ y_\lambda$ values as given in the last column of Table 1, which are the values that we used in our analysis.

%\begin{center}
%\begin{table*}[h]
\begin{table}
\centering
\begin{minipage}[t]{6cm}
\caption{Coefficients $x_\lambda$ and $y_\lambda$ for extinction and disk contributions\label{tab1}}
%\centering    
\begin{tabular}{lccc}
%\tablewidth{\columnwidth}
\hline\hline \\
{Filter} & {$x_\lambda$\footnote{From \citet{1985ApJ...288..618R}}} &{$y_\lambda$} & {$y_{\lambda}$\footnote{Using $y_\lambda \propto \lambda^3$.}} \\ 
\hline \\
%\startdata
%checar valores tabla  XXXXXXXX
I       &  4.82 & 0.064  & 0.08 \\
J	&  2.82 & 0.15\footnote{From \citet{2007RMxAA..43..155L}}	& 0.15   \\
H	&  1.75 & 0.40$^{c}$ 	& 0.43   \\
K	&  1.12 & 1.19$^{c}$ 	& 1.12   \\
L      &  0.58 &  4.86   & 4.93 \\
\end{tabular}
\end{minipage}
\end{table}
%\end{center}
%\tablenotetext{1}{From \citet{1985ApJ...288..618R}}
%\tablenotetext{2}{This paper.}
%\tablenotetext{3}{From \citet{2007RMxAA..43..155L}}.
%\end{deluxetable}
%\end{center}

As is shown in LS07 stellar masses can be obtained from the solutions of the minimization of the quadratic error $Err$  given by
\begin{equation}
Err^2 \,=\, {\sum_{\lambda=1}^{n}}  \frac {\left(m_\lambda \,-\, m_\lambda^{obs}\right)^2}  {n}
\end{equation}
where $m_\lambda^{obs}$ is the observed magnitude at each wavelength $\lambda$ and  $m_\lambda$ is the magnitude that should be observed according to the model:
\begin{equation}
m_\lambda \,=\,  m_\lambda^0\left(\mathrm{mass,age}\right)  \,+\, d \,+\,\alpha x_\lambda \,-\,  \beta y_\lambda.
\end{equation}
In eq. (4) the stellar magnitude is the sum of the absolute magnitude for a certain mass and age, plus the distance modulus $d$ to the object, plus the extinction correction and the infrared excess contribution from the circumstellar disk.  The error is then minimized with respect to its four parameters:  mass, age, $\alpha$ and $\beta$.  This minimization procedure is a set of linear equations for $\alpha$ and $\beta$. To deal with mass and age we compute $Err$  for each single mass and age taken from pre-main sequence evolutionary models. Then one seeks the minimum of $Err$ consistent with the additional constraints  $\alpha >$ 0 and $\beta >$ 0 obtained for each star.  This method is called the extinction-disk principal vector, XDPV method.

To test the XDPV method we  will use only the PMS tracks from \citet{1997MmSAI..68..807D} and \url{http://www.mporzio.astro.it/~dantona/prems.html} that provide luminosities and temperatures for pre-main sequence stars in a wide low-mass range, from 0.017 to 3 $M_{\sun}$ and ages from $10^4$ to $10^8$ yr.  The use of other evolutionary tracks, \citep[e.g.][]{1999ApJ...525..772P}, has been discussed in {LS07}. These authors argue that these tracks are found to produce similar mass results within 20\%. Luminosity and effective temperature of the evolutionary tracks are converted into absolute magnitudes using bolometric corrections and normal colors given in \citet{1995ApJS..101..117K}.

\section{Masses of young stellar cluster stars}

In LS07 we showed that if the masses of young stars are known, e.g. spectroscopically, the proposed XDPV method is a powerful tool to derive their ages or conversely, if the ages are known the masses can be extracted. However, both cannot be derived simultaneously.  This is due, as pointed out in LS07, to the fact that the minimal of $Err$ as a function of mass and age is a region that resembles a long and narrow canyon (c.f. their Fig. 7) that spans a wide range in masses and ages, providing a continuum of possible solutions. It is then necessary to specify an age for the PMS object to determine its mass.

However, when dealing with a cluster of stars with similar ages, a representative age of the cluster may be used for all the individual members.  Doing so requires a compromise age that would compensate the errors in mass by assuming a too young age for some objects  with those errors derived  from assuming an older age. The median age bisects the age histogram  in equal parts, so that an equivalent number of members
is either younger or older. Furthermore, the median is a robust indicator
of the central tendency, its value is the same whether one uses the histogram of ages
in linear or logarithmic values, and in general is a better choice when we only have
one number to specify a distribution.  For these reasons, we have chosen to use the median age of each cluster. 

We will show that this age selection gives consistent results for the resulting mass histograms of the well studied young stellar clusters of  {Taurus-Auriga} and  {Orion,} through the comparison of the IMF obtained by other authors using spectroscopic or photometric methods, and the IMF obtained applying our algorithm. 
We note that  in the young stellar clusters studied here, the median age 
has a lower value than the mean, due to a fraction of younger stars in the tail of
the distribution. For  {Taurus-Auriga} we obtain $\log \rm{median}\,=\,$5.8 (0.63 Myr) and  $\log \rm{mean}\,=\,$5.95 (0.89 Myr), while for  {Orion} $\log \rm{median}\,=\,$5.6 (0.4 Myr) and $\log \rm{mean}\,=\,$5.8 (0.63 Myr), as is described in detail below.
These values are significantly younger than what is commonly assumed as the age of the clusters.  But this has to be that way, because one usually refers to the epoch at which star formation began, which is the upper limit of the age distribution.

The method produces very good results as is shown below. Although it may seem a disadvantage to have to specify a cluster age, it is still far less information that the requirement of spectral types in spectroscopic methods, or a parametric description of the age distribution required in some other photometric methods.

\section{ {Taurus-Auriga}}

As pointed out by \citet{1995ApJS..101..117K}, hereon KH95, the  {Taurus-Auriga} is an ideal laboratory to study low-mass star formation, dominated by low-mass stars with little extinction, and so has been the subject of many investigations.
The list of known members of the region has grown through the years \citep{1979ApJS...41..743C,1988cels.book.....H,1995yCat.5073....0H,
1995ApJS..101..117K,
2002ApJ...580..317B,2004ApJ...617.1216L,2006ApJ...647.1180L,2006A&A...446..485G,2007A&A...468..405S}, with an increasing emphasis towards completeness.

As we discussed above, to estimate masses the XDPV method requires age estimates in addition
to near-infrared photometry. We will take this age from the histogram of ages presented by KH95 (their Fig.~16).  They have derived these ages together with masses  from
a full set of visible and infrared photometry and spectral types for known  {Taurus-Auriga} members
in 1995. Their age distribution spans from $\log(\rm{age(yr)})\,\equiv\, \log T\,=$ 4 to 6.5 and we calculate a mean value of 5.95 and a median value of 5.8 ($T$=6.3$\times$10$^5$ yr),  which on one hand is smaller than the commonly assumed value of 6.3, and also smaller than the mean. 
We may go on to compare masses derived by our method.  KH97 present spectral types and effective
temperatures for 139 stars. They use 103 of them to construct a mass histogram
from H-R diagrams with \citet{1994ApJS...90..467D} tracks with CMA opacities.  We took 189 stars with
$JHK$ photometry from their list, and calculated masses as described in Sect.~2 above, using the median
age derived  ($\log T$=5.8) from KH95,  and  the evolutionary models of \citet{1997MmSAI..68..807D} and \url{http://www.mporzio.astro.it/~dantona/prems.html}.

\begin{figure}
\includegraphics[width=\columnwidth]{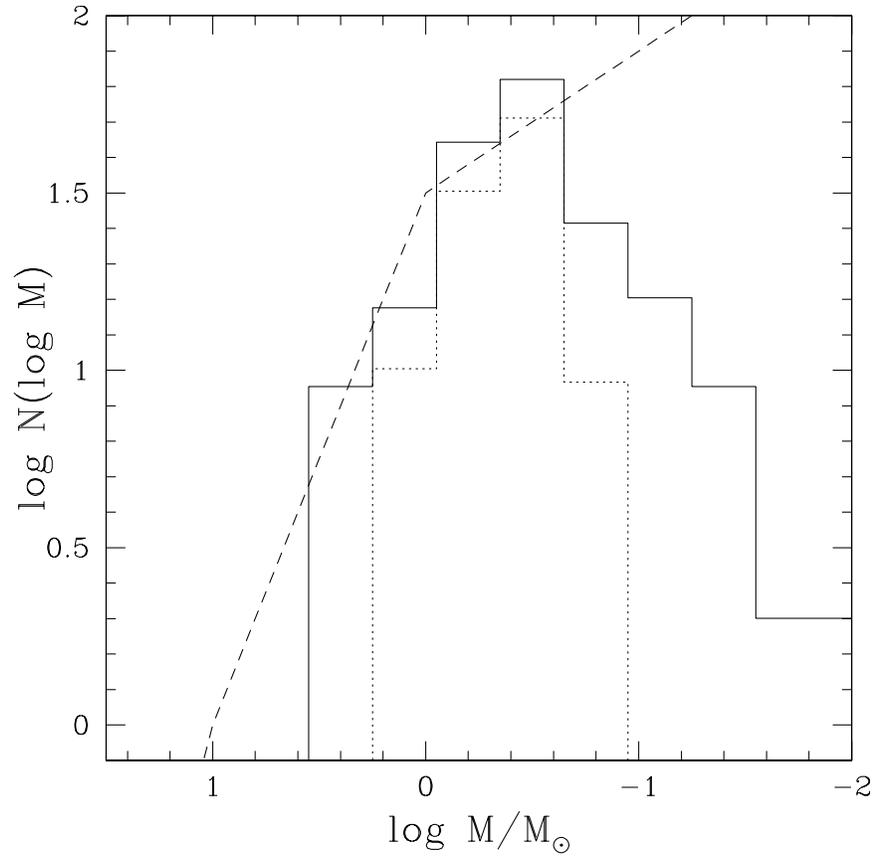}
\caption{IMF in  {Taurus-Auriga} from {\citet{1995ApJS..101..117K}} data processed by the XDPV method (shown as solid lines) compared to {\citet{1995ApJS..101..117K}} mass histograms (dotted lines).}
\label{fig2}
\end{figure}

We compare mass histograms in Fig.~\ref{fig2}. To do so, we have
re-binned KH95's histogram in logarithmic 0.3 dex bins, by assigning random masses 
to each star within its own bin, and then re-binning in the logarithmic bins. We repeated this
process 100 times to produce the mean histogram that is shown by dotted lines
in Fig.~\ref{fig2}, while our mass histogram is  shown as solid lines. For reference the Miller-Scalo IMF \citep{1979ApJS...41..513M} is shown as a dashed line.   The general shape from 0.3 to 1 $M_{\sun}$
is remarkably similar in both histograms and both peak at the same value where a 
turnover is observed. The excess number of
stars in our analysis (189 of 103) appear in three regions of the distribution. Some are in equal proportions
in the three bins from $\log M\,=\,$0.1 to -0.5 $M_{\sun}$, some lie in the high-mass end of the distribution which
nevertheless agrees quite well with the IMF of \citet{1979ApJS...41..513M}, and the majority are distributed in low-mass bins
$\log M/M_{\sun}\,<\,$-0.8, that most probably were too faint to provide a reliable spectral type by KH95.
However, this low-mass region is not unbiased or complete, as is pointed out in their paper.

\citet{2002ApJ...580..317B,2004ApJ...617.1216L} and \citet{2006ApJ...647.1180L} have paid attention to this fact, and have conducted 
spectroscopic surveys of several regions in the  {Taurus-Auriga} clouds in order to identify
all low-mass stars (many of which are brown dwarfs) to get complete unbiased samples 
of the IMF. This IMF is the result of photometry and spectroscopy to determine luminosities and effective temperatures, complemented with \citet{1998A&A...337..403B} evolutionary tracks to derive masses.  The more consolidated example of this IMF is given in \citet{2004ApJ...617.1216L}. 
We have taken 2-MASS $JHK$ data given in \citet{2006ApJ...647.1180L} for one of such complete regions that are
consistent with \citet{2004ApJ...617.1216L} IMF.  In this new article, Luhman incorporated some 20 newly discovered
brown dwarfs into a large region that encompasses about half of the known Taurus population,
and thus constitutes a significant sample. With our method we were able to obtain solutions for 125 objects out of the presented list of 156,  again
using the median age derived from KH95,  and  the evolutionary models of \citet{1997MmSAI..68..807D} and \url{http://www.mporzio.astro.it/~dantona/prems.html}.

\begin{figure}
\includegraphics[width=\columnwidth]{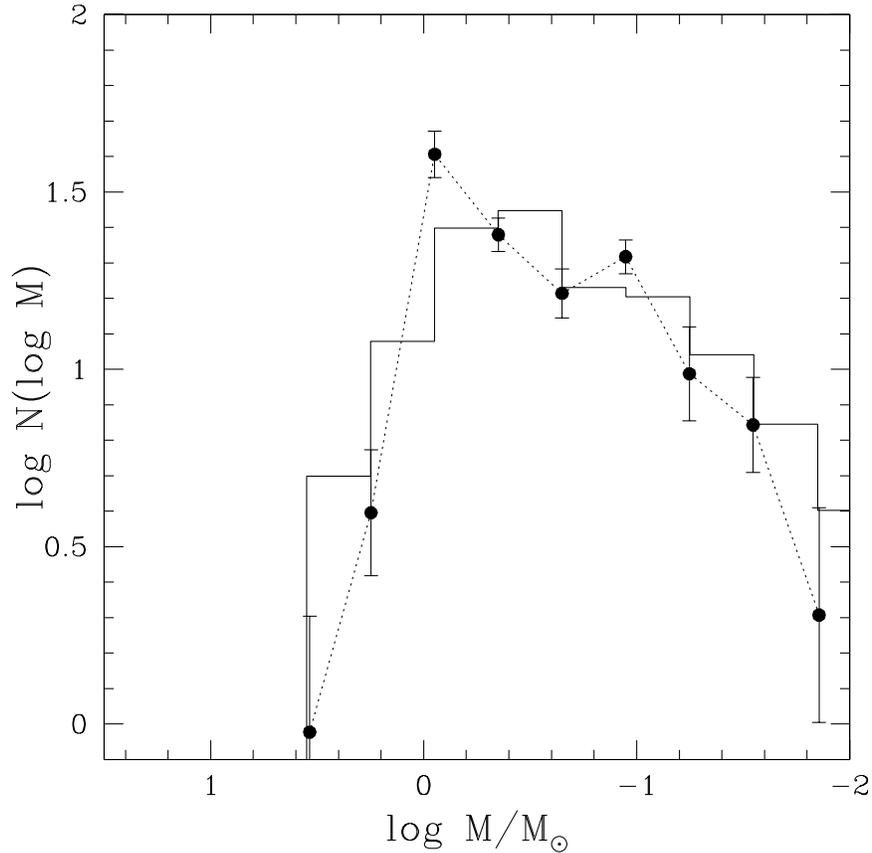}
\caption{IMF in  {Taurus-Auriga} from \citet{2006ApJ...647.1180L} data as obtained by the XDPV method (shown as solid lines) compared to {\citet{2004ApJ...617.1216L}}'s IMF (dotted lines).}
\label{fig3}
\end{figure}

As can be observed in  Fig.~\ref{fig3} the resemblance between the IMF obtained by \citet{2004ApJ...617.1216L} and the IMF obtained using the XDPV method for the {Taurus-Auriga} region is remarkable, with a probability of over 60\% of being random samples of the same population.  For a Kolmogorov-Smirnov test of this result and others, see section 6.

Furthermore, the XDPV solutions for both data sets \citep{1995ApJS..101..117K,2004ApJ...617.1216L} presented in Figures 2 and 3, resemble one another better that those presented in the original papers. We conclude that this agreement strengthens our results.

\section{ {Orion Nebula Cluster}}

\subsection{About the IMF}

The  { Orion Nebula Cluster} (ONC) centered on the Trapezium
OB stars is the richest of any nearby clusters and has been 
studied extensively. Numerous studies have targeted this
cluster to determine its underlying population and the associated IMF \citep[e.g.][]{1997AJ....113.1733H,2000ApJ...540.1016L,2002ApJ...573..366M}.  We
will compare our own results to those of two important studies, one spectroscopic
and one based on near-infrared photometry. 

\citet{1997AJ....113.1733H} has obtained spectral types for 934 visible stars.  This
information, supplemented by optical photometry, allowed her to populate an
H-R diagram with pre-main sequence evolutionary tracks, and to extract mass and age information.
She notes, however, that  large uncertainties arise from the choice of a particular
set of evolutionary tracks \citep[see also][]{2004ApJ...604..741H}. 
Nevertheless,  \citet{1997AJ....113.1733H} chooses \citet{1994ApJS...90..467D} evolutionary models to display the
ONC's IMF (her Fig.~17), and it has become a seminal reference for this region. 
The derived distribution of stellar ages of the ONC population conform a 
distribution that spans a wide range of ages.  It starts at $\log T\,=\,$3.5 (3000 yr) and increases
gradually until $\log T\,=\,$6.3 (2 Myr), then decreases abruptly and continues at a low
constant pace up to $\log T\,=\,$7.8 (63 Myr).  This distribution has a mean value of $\log <T>\,=\,$5.84 (0.7 Myr) although some authors quote 0.8 Myr.  It has also been represented
by a constant rate from $\log T\,=\,$5 (0.1 Myr) to $\log T\,=\,$6 (1 Myr). In our treatment  of the OMC cluster we will choose
the median age, $\log T\,=\,$5.6 (0.4 Myr), which we believe is the best compromise as is discussed above for Taurus-Auriga.
This is in agreement with \citet{2000ApJ...540.1016L} work that also quotes a median age of 0.4 Myr.

\citet{2000ApJ...533..358M} developed a Monte Carlo method to model the IMF based in obtaining the KLF
from a series of probability distributions: extinction, infrared excess, age and the 
IMF modeled as a series of power laws.  They applied this method to $JHK$ observations of the ONC in \citet{2002ApJ...573..366M}.
The age distribution was chosen as a uniform distribution from 0.2$\times$10$^6$ to 1.4$\times$10$^6$ yr. 
The extinction distribution was derived from $(J-H)$ vs. $(H-K)$ diagram by de-reddening
sources down to the classical T Tauri star locus of \citet{1997AJ....114..288M}.
After this, the infrared excess distribution was obtained from the remaining excess in the $(H-K)$
color after subtracting a histogram of $(H-K)$ colors of field stars, and this excess color was assumed to
arise exclusively from excess disk emission at $K$.
From an original set of $\sim$1000 sources, they took an
extinction limited sample ($A_V\,<\,$17) of 583 stars, which is said to be complete down to 0.017 $M_{\sun}$. 

\begin{figure}
\includegraphics[width=\columnwidth]{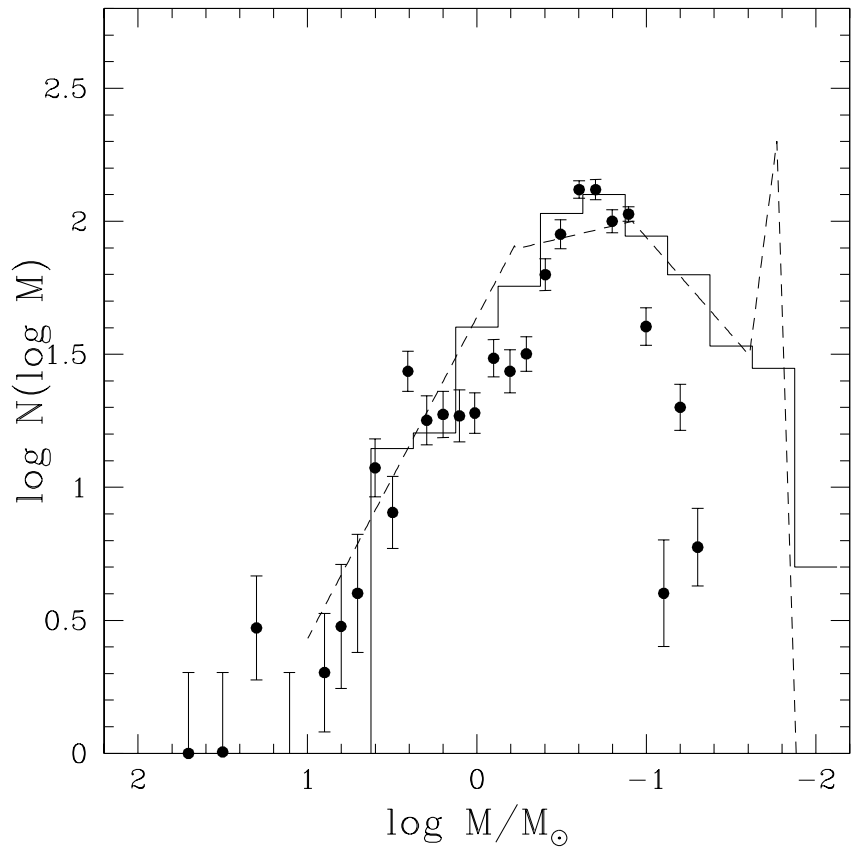}
\caption{ONC's IMF (solid histogram) compared to  {\citet{1997AJ....113.1733H}} (points with error bars)
and {\citet{2002ApJ...573..366M}} (dashed line).}
\label{fig4}
\end{figure}

We took $JHK$ photometry from the \citet{2002ApJ...573..366M} published list, assumed a median age of $\log T\,=\,$5.6 from \citet{1997AJ....113.1733H} age histogram, used \citet{1997MmSAI..68..807D} and \url{http://www.mporzio.astro.it/~dantona/prems.html} evolutionary tracks, and assumed a distance of 400 pc \citep{2002ApJ...573..366M}. 
With these ingredients we
applied our XDPV method to the 699 object where no confusion flags are found
and photometry is available in all $JHK$ filters.  We were able to obtain solutions consistent with
both $\alpha\,>\,$0 and $\beta\,>\,$0 and an acceptable $Err\,<\,$0.3 mag for 612 of  these stars.
Objects for which no solution was found are, for example, stars more massive than those present in  \citet{1997MmSAI..68..807D} and \url{http://www.mporzio.astro.it/~dantona/prems.html} evolutionary tracks, maximum of 3 $M_{\sun}$, which includes all the  {Trapezium} stars,  {BN} and  {$\Theta^2$Ori A}. From the 612 PMS objects
we then selected  578 with $A_V\,<\,$17 to display in Fig.~\ref{fig4}. This  compares our IMF (solid histogram) with that of \citet{1997AJ....113.1733H}, points with error bars, and
\citet{2002ApJ...573..366M} shown as a dashed line.  The agreement with \citet{2002ApJ...573..366M} is excellent, that is, the
initial slope for high-mass stars, the flattening and the position of the turnover, followed by the negative slope in the subsolar mass spectrum all agree quite well. The only  exception is for the low-mass
secondary peak in the substellar region ($\log M/M_{\sun}\,=\,$ -1.8).  Unfortunately, those objects come from the low-brightness peaks in the $H$ and $K$ histograms in \citet{2002ApJ...573..366M} but are too faint and red, and therefore, absent  in $J$.  Consequently, they do not appear in our data.  

The agreement with the \citet{1997AJ....113.1733H} IMF is quite good in the  -0.9 $\,< \log M/M_{\sun} <\,$ 0.45  range.  Massive stars (M$>$3  M$_{\sun}$) are missing in
our histogram as mentioned above, due to the limitted mass range in
the evolutionary tracks. In the  $\log (M/M_{\sun}) \,<\,$ -1 range, the \citet{1997AJ....113.1733H} survey is most likely
incomplete for sources with $A_V\,>\,$2.5 \citep{2000ApJ...540..236H}. The position of the turnover also agrees, although the exact position may be at slightly lower masses as has been revised in \citet{2000ApJ...540..236H} by using updated evolutionary tracks and transformations.

We conclude that the IMF we obtained with the XDPV method is a robust representation of that obtained by other methods. For a more through comparison of these distributions see section 6.

\subsection{Extinction and infrared excess}

\begin{figure*}
\includegraphics[width=\textwidth]{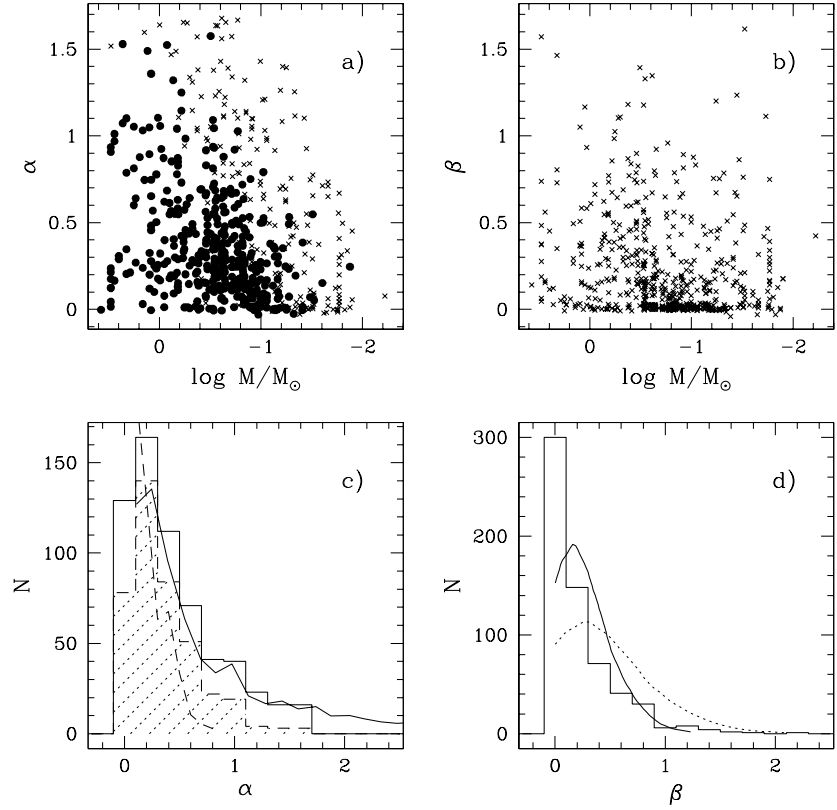}
\caption{ Top panels:
 $\alpha$ and $\beta$ as function of mass.  In panel a)  sources also
in the survey of  \citet{1997AJ....113.1733H} are marked as dots. Bottom panels: histograms of $\alpha$ and $\beta$ in solid lines, are
compared to c) the extinction probability distribution of \citet{2002ApJ...573..366M} solid line,
and the extinction data in \citet{1997AJ....113.1733H} in dashed line to its corresponding 
dashed line histogram. d) The infrared excess probability distribution of \citet{2002ApJ...573..366M}
if considered as $E(H-K)$ (dashed line) or $K$-excess (solid line). }
\label{fig8}
\end{figure*}

In addition to the mass data, our method also gives information about the 
extinction $\alpha$ and infrared excess $\beta$ of the sources.  We display
these two quantities as histograms in Fig.~\ref{fig8}, and compare them with extinctions and
excesses used in \citet{2002ApJ...573..366M} and \citet{1997AJ....113.1733H}.

The bottom left panel shows the histogram of extinction $\alpha$ as a solid histogram.  
This quantity can
be directly understood in terms of the visual extinction,  $\alpha \,=\, A_V/10$.  It is then easy to compare it with
the extinction probability distribution presented in \citet{2002ApJ...573..366M}, shown here as a solid line, which is very
close to our result. In order to compare with the spectroscopically derived 
extinctions in \citet{1997AJ....113.1733H}, we show the dashed histogram for those sources 
that were analyzed by us and that are also part of the \citet{1997AJ....113.1733H} survey, and  compare it
to the dashed line obtained from extinctions derived by her.  A discrepancy is notorious
for the very low extinction ($A_V \,<\,$ 2) sources, where \citet{1997AJ....113.1733H} finds 
most sources and our histogram turns over.

In the bottom right panel we show the histogram for $\beta$ compared to the infrared
excess distribution in \citet{2002ApJ...573..366M}. However, there are two possibilities for this comparison. 
If the abscissa in \citet{2002ApJ...573..366M} Fig.~8b is the excess in $H-K$, then from eq.~(2) $\beta\,=\, E_\mathrm{H-K} / (y_\mathrm{K}-y_\mathrm{H})$,
while if the abscissa is taken as the excess in $K$ alone, then $\beta \,=\, -E_\mathrm{K} / y_\mathrm{K}$, and
given the values in Table~1, they are not equal.  We show both possibilities 
in Fig.~5d. The infrared excess  is best represented by $\beta$ and we find a better agreement when the infrared excess distribution is represented by $-E_\mathrm{K}$, rather than $E_\mathrm{H-K}$. 

In the top panels of Fig.~\ref{fig8} we show $\alpha$ and $\beta$ as functions of mass $\log M$. As a general rule, no dependence of these two quantities with mass is found, as  $\alpha$ and $\beta$ acquire all their values for any mass in the range from 0.017 to 3.0 $M_{\sun}$.  The one exception is observed in the case of $\alpha$ (Fig.~5a): there are no
low-mass stars at high extinctions. This is more notorious in the case 
of those stars also observed by \citet{1997AJ....113.1733H}, marked as solid dots, where 
a diagonal line in the upper rightmost part of the diagram can easily be drawn.
This is expected to be the case, since low-mass stars cannot be observed at high extinctions for given observation times, and since this fact was not introduced a priori in the method, it constitutes another confirming result of the XDPV approach. In this diagram it is then possible to extract an extinction limited sample, as a rectangular box that lies below the diagonal line corresponding to each sample. It can be seen that
\citet{1997AJ....113.1733H} survey is unbiased for $\log {M/M_{\sun}} \,>\,$-1.1 and $A_V \,<\,$3, 
in close agreement with her findings \citep[see also][]{2000ApJ...540..236H}. On the other
hand, the completeness limit in all three $JHK$ filters in the
\citet{2002ApJ...573..366M} survey would go up to $\log {M/M_{\sun}} \,>\,$-1.8 and $A_V \,<\,$6. Finally. given that the infrared excess $\beta$ is due to the disk contribution we find that 300 stars out of 612 (close to 50\%) presumably possess associated disks ($\beta>0$), independently of their mass.  This is consistent with recent Spitzer studies that show that the percentage of low-mass PMS stars with disk in ONC is about 50\% \citep{{2006ApJ...646..297R},{2007ApJ...671..605C}}

\section{Statistical significance of IMFs}

For two young stellar clusters a general concern is whether the individual IMFs are different or not, given that the shapes of the IMFs look similar. To quantify this we performed a Kolmogorov-Smirnov (K-S) test.   First, we compared the mostly accepted mass distributions for Taurus and Orion, to show that they are significantly different.  Second, we compared the IMFs obtained by the XDPV method to both the Taurus and Orion cases.

To perform the K-S test we developed a routine that generates a random population of stellar masses from any given mass distribution $N(\log M)$ using the well known accept-reject algorithm.  For each distribution that we tested, we generated a number of random stellar masses equal to the number of stars that were used to obtain the original distribution, in order to keep the same stochastic variability that would be expected in a different realization of the distribution.
For each comparison between different distributions, we took 100 realizations of each distribution being compared.  and computed the two distribution K-S probabilities that the samples are drawn from the same parent distributions (null hypothesis). We report the mean of these 100 comparisons (K-S MP).

We first tested the IMFs reported by \citet{2004ApJ...617.1216L} for Taurus-Auriga (hereon TL) to those of \citet{1997AJ....113.1733H} and \citet{2002ApJ...573..366M}, hereon OH and OM respectively, for Orion.  The OH distribution has the advantage of being derived spectroscopically (as well as the TL one), but is incomplete for masses lower than $\log(M/M_\sun) < -0.9 $ as we already mentioned.  For this reason we limited this distribution to the medium mass range in the comparisons. We therefore label it as OHM. Similarly as we already discussed, the OM distribution has a secondary peak at the very low mass end which we cannot test because the sources involved are below the J filter detection limit. Therefore the OM distribution is limited to $\log(M/M_\sun) > -1.6$ for all tests here.  The comparisons between TL and OHM yield a K-S mean probability (K-S MP) of 3$\times$10$^{-11}$ of being drawn from the same parent distribution, while TL and OM gives K-S MP of 1$\times$10$^{-4}$.  These very small values makes us confident that our test is consistent with previously known results (e.g. \citet{2008ApJ...672..410L}) that the distributions are indeed different.

We then compared the distribution of XDPV masses obtained for the Taurus-Auriga region (hereon TX) to TL, obtaining a K-S MP of 21\% therefore confirming the posibility of both distributions being equivalent representations.  The comparison of TX to OHM rules out the null hypothesis with a K-S MP of 6$\times$10$^{-5}$.  It should be mentioned however, that TX is not that different from OM (K-S MP of 1\%) mainly because in the low mass range $-1.6 < \log(M/M_\sun) < -0.9 $ TL is quite similar to OM (K-S MP = 67\%).  That is to say, the low mass slope of Taurus and Orion IMFs are similar.  Nevertheless, the agreement between TX and TL is sustained.

Finally, the distribution of XDPV masses obtained for the Orion region (hereon OX) compares with OHM with a K-S MP of 3\% in the medium mass range, and a somewhat better agreement is seen to OM with K-S MP = 9\% in the whole range of masses $\log(M/M_\sun) > -1.6$ .  On the other hand, OX and TL cannot be accepted as a match with a very low K-S MP of 2$\times$10$^{-7}$.  This proofs that OX is not consitent with the Taurus IMF but has a fair probability of being drawn from the same parent population of the Orion Nebula Cluster.

\section{Summary}

We used the extinction-disk-principal vectors approach, which is called the XDPV method,   reported previously in  \citet{2007RMxAA..43..155L} to show that it is a powerful tool to estimate masses of pre-main sequence stars in clusters.  The method requires a minimum of information, using as little as $JHK$ (or $IJK$ or $JKL$) near-infrared photometry, supplemented by a set of PMS evolutionary tracks \citep[e.g.][]{1997MmSAI..68..807D} and the median age of the cluster.  For each star in the cluster, we are able to estimate the contribution of the extinction vectors $\vec{X_\mathrm{cc}}$ in the color-color diagram and  $\vec{X_\mathrm{cm}}$ in the color-magnitude diagram, and the disk vectors $\vec{D_\mathrm{cc}}$ in the color-color diagram and $\vec{D_\mathrm{cm}}$ in the color-magnitude diagram, using \citet{2005RMxAA..41...61D} accretion disk models grid of spectral energy distributions.  The observed absolute magnitude at each wavelength $\lambda$ of a PMS object is obtained via $m_\lambda \,=\,  m_\lambda^0\left(\mathrm{mass,age}\right)  \,+\, d \,+\,\alpha x_\lambda \,-\,  \beta y_\lambda$, where the first term corresponds to the absolute magnitude of the naked object for a certain mass and age, followed by the distance modulus $d$, the extinction correction ($\alpha x_\lambda$) and the infrared excess contribution ($-\beta y_\lambda$) from the circumstellar disk.  
It is shown that if a representative age of the cluster, such as the median, is known, the masses of each individual star can be statistically obtained from  its near-infrared photometry alone. The XDPV method is tested in the well studied regions of  {Taurus-Auriga} and the {Orion Nebula Cluster} by extracting their Initial Mass Function.  These IMF are in excellent agreement (K-S test) to those given by \citet{1995ApJS..101..117K} and \citet{2004ApJ...617.1216L} for Taurus and \citet{1997AJ....113.1733H} and \citet{2002ApJ...573..366M} for Orion.
Since our algorithm also yields the extinction and disk contribution for each star the distributions can be obtained. 
The overall extinction distribution for the Orion cluster is analyzed and compares well with previous work, the comparison to \citet{2002ApJ...573..366M} shows that the parameter $\alpha$ for the extinction vector is  $-E_\mathrm{K}$ rather than $E_\mathrm{H-K}$. The frequency of PMS low-mass stars with disks, represented by the parameter $\beta$,  is about 50\% in the Orion sample.  It is also seen that the number of sources observed decreases for high values of extinction $\alpha$, confirming a well known and expected observational effect, and allowing us to draw a complete extinction limited sample a posteriori. We conclude that our XDPV algorithm can be applied to study the IMF of young stellar clusters, as well as the distributions of extinction and disk infrared excess.

\acknowledgments We are grateful to an anonymous referee for very valuable comments that aided in the final
version of the manuscript.

\bibliographystyle{rmaa}
\bibliography{discos09} % your references Yourfile

\end{document}